\documentclass[aps,preprint,showpacs,superscriptaddress,groupedaddress,prb,sort&compress]{revtex4-2}

\usepackage{graphicx}  
\usepackage{color}
\usepackage{subfigure}
\usepackage{amsmath}

\newcounter{JW}

\usepackage{hyperref}

\begin{document}

\title{The relation between the radii and the densities of magnetic skyrmions}

\author{Yu-Jiao Bo}
\email{bo1323650389@163.com}
\author{Wen-Wen Li}
\email{lww1514148@163.com}
\author{Yu-Chen Guo}
\email{ycguo@lnnu.edu.cn}
\author{Ji-Chong Yang}
\email{yangjichong@lnnu.edu.cn}
\affiliation{Department of Physics, Liaoning Normal University, Dalian 116029, China}

\begin{abstract}
Compared with the traditional magnetic bubble, skyrmion has smaller size, better stability therefore is considered as a very promising candidate for future memory devices.
When skyrmions are manipulated, erased and created, the density of skyrmions can be varied, however the relationship between the radii and the densities of skyrmions needs more exploration.
In this paper, we study this problem both theoretically and by using the lattice simulation.
The average radius of skyrmions as a function of material parameters, the strength of external magnetic field and the density of skyrmions is obtained and verified.
With this explicit function, the skyrmion radius can be easily predicted, which is helpful for the future study of skyrmion memory devices.
\end{abstract}



\maketitle
\section{\label{level1}Introduction}

Skyrmion is a topological soliton originally proposed to describe the baryons~\cite{Skyrme:1962vh}.
In condensed matter, a particle-like object known as magnetic skyrmion was introduced theoretically in 1989~\cite{magneticSkyrmionIntro}.
It was observed for the first time in 2D magnetic systems~\cite{discover1,discover2,discover3,discover4} involving Dzyaloshinskii-Moriya interactions~(DMI)~\cite{dmi1,dmi2}.
Compared with the traditional magnetic bubble, the skyrmion is smaller, more stable and needs lower power to manipulate, therefore, it has been proposed that the skyrmion is a promising candidate for high density, high stability, high speed, high storage and low energy consumption memory devices~\cite{radius1,lowconsumption,review1}.
As a result, the magnetic skyrmions have draw a lot of attention and are studied intensively recently~\cite{review1,review2,review3,reviewA,reviewB}.

A prerequisite for the use of skyrmions in devices is the knowledge of the relationship between the size of a skyrmion and parameters such as exchange strength, DMI strength and the strength of external magnetic field.
Such a relationship can be investigated by solving the Euler-Lagrange equation of a skyrmion, for example numerically~\cite{radius5} or by using an ansatz~\cite{radius1}, or by using the harmonic oscillation expansion~\cite{radius2}, or by an asymptotic matching~\cite{radius3}.
It has been noticed that the radius of a skyrmion in the skyrmion phase is much smaller than that of an isolated skyrmion~\cite{radius2}.
Both the radii of an isolated skyrmion and the skyrmions in the skyrmion lattice were studied quantitatively in Ref.~\cite{radius4}.
In particular, numerical results were obtained for the equilibrium radii of skyrmion lattices.

However, as a potential candidate for storage, the skyrmion is meant to be manipulated, erased and created.
In this case, the number of skyrmions can vary from just only one to filling the entire skyrmion lattice.
The transformation of a skyrmion lattice to the saturated state is continuous, in this process, the skyrmion lattice gradually decomposes into isolated skyrmions in the saturated state~\cite{reviewB}.
In this paper, we study the average radius of skyrmions with the density of skyrmions in the range between a single isolated skyrmion and the skyrmion lattice.
While the results have been obtained for a single isolated skyrmion, and for skyrmion lattices, up to our knowledge, the radius of a skyrmion when the density of the skyrmions is between the skyrmion lattice and the single isolated skyrmion is poorly understood at a quantitative level.

The rest of paper is organized as the following.
The analytical and numerical results based on circular cell approximation are established in Sec.~\ref{level2}.
In Sec.~\ref{level3}, we introduce the lattice simulation of Landau-Lifshitz-Gilbert~(LLG) equation.
We compare the theoretical results with lattice simulation in Sec.~\ref{level4}.
A summary is made in Sec.~\ref{level5}.

\section{\label{level2}Circular cell approximation}

The local magnetic moment of a skyrmion can be parameterized as
\begin{equation}
{\bf n}(r,\phi,z)=\sin [m\theta (r)+\gamma] {\bf e}_{\phi}+g\cos [\theta (r)] {\bf e}_z,\\
\label{eq.2.1}
\end{equation}
where $r, \phi, z$ are coordinates in a cylindrical coordinate, $\gamma$ is the helicity angle, $m= 1$ for a skyrmion  and  $m= -1$ for an anti-skyrmion, $g=\pm 1$.
The skyrmion number is $Q=-mg$.
In the following, we only consider the skyrmion with $Q=1$~($m=1,g=-1$).

By using the circular cell approximation, the skyrmions are viewed as sitting in circular cells with radius $R$, which means the boundary condition $\theta(0)=\pi$ and $\theta(R)=0$~\cite{radius4}.
In principle, $\theta(r)$ can be expanded using any Hilbert space.
Since the wave-function of ground state of the harmonic oscillator and the numerical solution of the Euler-Lagrange equation of a skyrmion are close in shape~\cite{radius2}, we use the Hilbert space of harmonic oscillator to expand $\theta (r)$.
We do not require $\theta'(r)=0$ as in Ref.~\cite{radius2} because $\theta'(r)\neq 0$ is allowed by the Euler-Lagrange equation, therefore the eigen-functions of odd energy levels are also included.
To impose the boundary conditions $\theta(0)=\pi$ and $\theta(R)=0$, $\theta(r)$ to the next-to-next-to leading order can be written as
\begin{equation}
\theta(r)=\sum _{n=0}^{2}C_n\phi _n(r)=\pi e^{-\frac{\omega r^2}{2}}-\frac{1+c R^2}{R}r \pi e^{-\frac{\omega r^2}{2}}+ c \pi  r^2 e^{-\frac{\omega r^2}{2}},\\
\label{eq.2.2}
\end{equation}
where $\phi _n$ are eigen-functions of harmonic oscillator, $\omega$ and $c$ are parameters to be determined.
By assuming the coefficients of the higher order terms are small, the power counting yields $c\sim 1/R^2$ and $1/R\ll 1$.

We concentrate on the case when the anisotropy is absent, the energy to be minimized is $F=2\pi \int _0^R dr r \mathcal{F}(r)$ with the energy density
\begin{equation}
\mathcal{F}(r)=2J\left\{\left[\left(\frac{1}{2}\frac{\partial \theta}{\partial r}+\frac{d}{2}\right)^2-\left(\frac{d}{2}\right)^2+\frac{\sin^2(\theta)}{4r^2}+\frac{d\sin(2\theta)}{4r}\right]-\frac{b}{2}(\cos (\theta)-1)\right\},
\label{eq.2.3}
\end{equation}
where $d\equiv D/J$ and $b\equiv B/J$, $J$ is the strength of local ferromagnetic exchange, $D$ is the strength of DMI, $B$ is the strength of the external magnetic field which is assumed to be parallel to the ${\bf z}$-axis.
For simplicity, we consider dimensionless parameters, the matching is discussed in Sec.~\ref{level4.4}.

Denoting $s\equiv 1/R$, $F$ can be expanded as $F=\hat{F}+\mathcal{O}(s^5)$ with
\begin{equation}
\begin{split}
&\hat{F}=-\frac{1}{72 s^4 \omega^3}\left\{-36 \pi ^3 b c^2 s^4 f_{1,2,1,0}+12 \pi ^3 b s \sqrt{\omega} \left(c+s^2\right) \left(6 c s^2 f_{1,\frac{3}{2},1,0}-\pi  f_{\frac{3}{2},\frac{3}{2},0,1} \left(c+s^2\right)^2\right)\right.\\
&\left.+3 \pi ^5 b f_{2,2,1,0} \left(c+s^2\right)^4+36 \pi ^4 b c s^2 f_{\frac{3}{2},2,0,1} \left(c+s^2\right)^2+72 \pi ^2 b s^3 \omega^{3/2} f_{\frac{1}{2},\frac{1}{2},0,1} \left(c+s^2\right)\right.\\
&\left.+144 \pi ^3 c^2 d s^4 \sqrt{\omega} f_{1,\frac{3}{2},1,1}-144 \pi ^2 c d s^4 \omega^{3/2} f_{\frac{1}{2},\frac{1}{2},2,0}+144 \pi ^3 d s^2 \omega^{3/2} f_{1,\frac{1}{2},1,1} \left(c+s^2\right)^2\right.\\
&\left.-48 \pi ^5 d \sqrt{\omega} f_{2,\frac{3}{2},1,1} \left(c+s^2\right)^4+288 \pi ^4 c d s^2 \sqrt{\omega} f_{\frac{3}{2},\frac{3}{2},2,0} \left(c+s^2\right)^2\right.\\
&\left.-24 \pi ^3 s \omega^{3/2} \left(c+s^2\right) \left(3 c s^2 \left(f_{1,\frac{1}{2},0,2}-f_{1,\frac{1}{2},2,0}\right)+2 \pi  f_{\frac{3}{2},\frac{1}{2},1,1} \left(c+s^2\right)^2\right)\right.\\
&\left.+72 \pi ^2 s^3 \omega^{5/2} f_{\frac{1}{2},-\frac{1}{2},1,1} \left(c+s^2\right)-72 \pi  d s^4 \omega^{5/2} f_{0,-\frac{1}{2},1,1}-36 \pi  s^4 \omega^3 f_{0,-1,0,2}\right.\\
&\left.+36 s^2 \omega \left[4 \pi  b \left(c^2 (-\text{Ci}(\pi )+\gamma _E -2+\log (\pi ))-4 c s^2+s^4 (-\text{Ci}(\pi )+\gamma _E-2+\log (\pi ))\right)\right.\right.\\
&\left.\left.+4 \pi ^{5/2} \sqrt{2} c d s^2 \sqrt{\omega}-\pi ^3 w \left(c+s^2\right)^2\right]-144 \pi  b s^4 \omega^2 (-\text{Ci}(\pi )+\gamma _E +\log (\pi ))\right.\\
&\left.+18 \pi  \omega \left(c^4+s^8\right) (-\text{Ci}(2 \pi )+\gamma _E+\log (2 \pi ))+3 \pi  s \omega \left[16 d \text{Si}(2 \pi ) \left(c^3+s^6\right)+16 \pi  d \left(c+s^2\right)^3\right.\right.\\
&\left.\left.+15 \pi ^{5/2} c s^2 \sqrt{\omega} \left(c+s^2\right)\right]-12 \pi ^3 \gamma \left(c^4+4 c^3 s^2+12 c^2 s^4+4 c s^6+s^8\right)\right.\\
&\left.-16 \pi ^{9/2} \sqrt{6} c d s^2 \sqrt{\omega} \left(c+s^2\right)^2+18 \pi ^2 s^3 \omega^2 \left(c+s^2\right) \left(\pi ^{3/2} \sqrt{\omega}-8 d\right)\right.\\
&\left.+72 \pi ^{5/2} \sqrt{2} d s^4 \omega^{5/2}-36 \pi ^3 s^4 \omega^3\right\},\\
\end{split}
\label{eq.2.4}
\end{equation}
where $\gamma _E$ is the Euler constant, ${\rm Ci}$ and ${\rm Si}$ are cosine and sine integral functions, and $f_{m,n,n_c,n_s}$ are constant numbers defined as
\begin{equation}
f_{m,n,n_c,n_s}\equiv \int _0^{\infty}dx e^{-m x}x^n\cos ^{n_c}\left(\pi e^{-\frac{x^2}{2}}\right)\sin ^{n_s}\left(\pi e^{-\frac{x^2}{2}}\right).
\label{eq.2.5}
\end{equation}
This seemingly lengthy expression of $\hat{F}$ is nothing more than a polynomial of $\sqrt{\omega}$ and $c$.
To achieve a higher precision, in principle, both the expansions of $\theta (r)$ and $F$ can be worked out for higher orders.

\begin{figure}[!htbp] \centering
\includegraphics[width=0.49\textwidth]{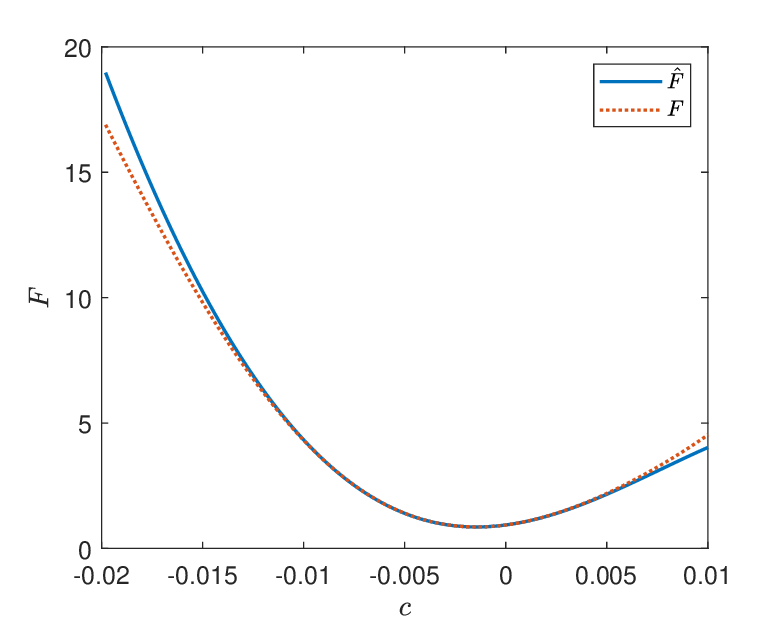}
\includegraphics[width=0.49\textwidth]{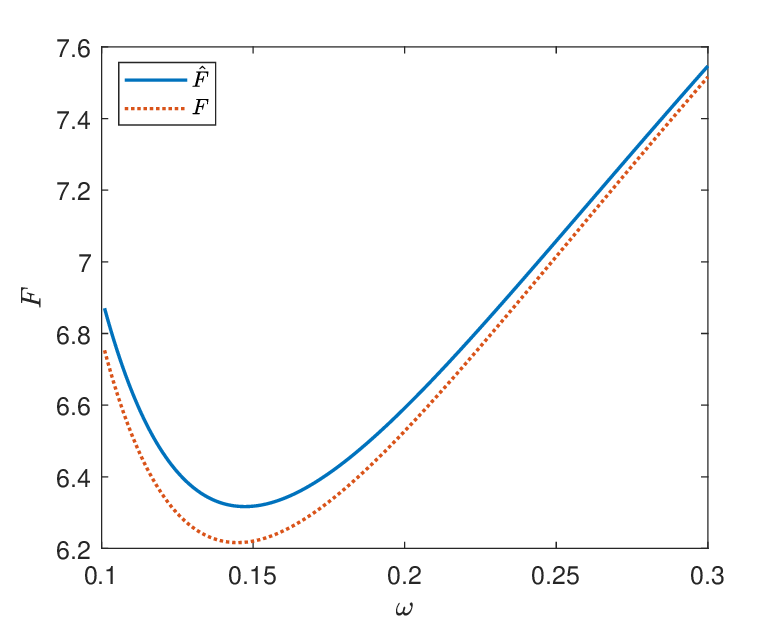}
\caption{\label{fig.freeenery}Compare $F$ with $\hat{F}$ at $d=0.4$, $b=0.1$, $R=20$. The left panel is $F$ and $\hat{F}$ at $\omega=0.15$ as functions of $c$, the right panel is $F$ and $\hat{F}$ at $c=-0.012$ as functions of $\omega$. }
\end{figure}

For $d=0.4$, $b=0.1$, $R=20$, in the region that $\omega \sim 0.15$ and $c\sim -0.01$, we compare $F$ with $\hat{F}$ in Fig.~\ref{fig.freeenery}. $\hat{F}$ can approximate $F$ well in the region concerned.
Especially, the positions where $F$ and $\hat{F}$ are minimized fit each other very well.
To minimize $F$, we use variational method, so that there are two equations $\partial F / \partial \omega = 0$ and $\partial F / \partial c = 0$, by which $\omega$ and $c$ can be solved.

By setting a threshold $h$ such that the sites with $n_z<h$ are determined as inside a skyrmion, the radius of a skyrmion can be obtained by solving the equation $\cos(\theta(r))=h$.
Considering the leading order approximation which correspond to $c=s=0$~(denoted as $\theta _{\rm LO}$), by solving $\cos(\theta_{\rm LO}(r))=h$, the radius of a skyrmion~(denoted as $r_s$) is approximately
\begin{equation}
r_s=\sqrt{\frac{2}{\omega}\log \left(\frac{\pi }{\cos^{-1}(h)}\right)},
\label{eq.2.6}
\end{equation}
where $\omega$ can be solved by the equations $\partial \hat{F} / \partial \omega = \partial \hat{F} / \partial c = 0$ mentioned above.
We choose the solution that $\omega $ and $c$ are real numbers, and $|c|\ll 1$.

With the numerical solutions of $\omega$, we can investigate the change of $r_s$ as function of $s$.
For this purpose, we define $\hat{r}=r_s/r_{\rm iso}$ where $r_s$ is calculated by $\omega$ solved at $s$, and $r_{\rm iso}$ is the radius of a single isolated skyrmion which is calculated by $\omega$ solved at $s\to 0$.

\begin{figure}[!htbp] \centering
\includegraphics[width=0.75\textwidth]{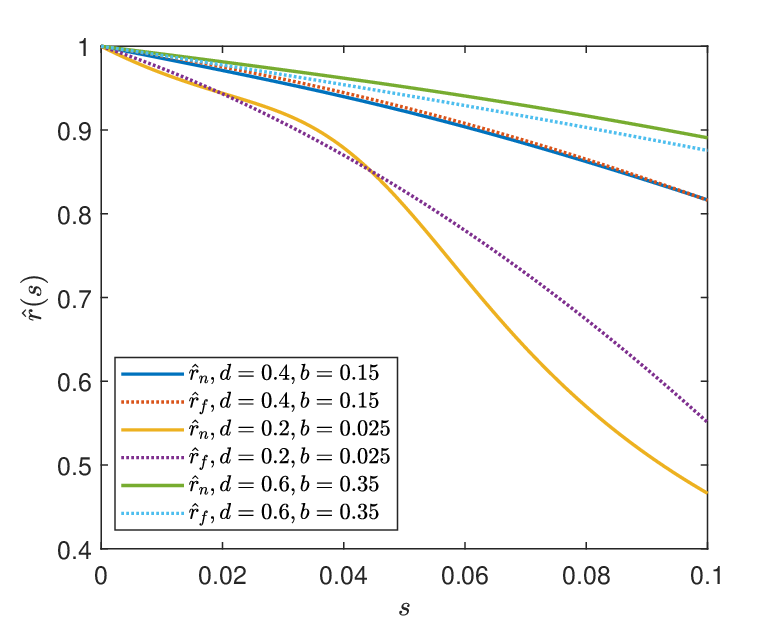}
\caption{\label{fig.fitrhat}$\hat{r}_f$ Compared with $\hat{r}_n$. }
\end{figure}

Since the numerical solutions are inconvenient to use, we fit the solutions as bilinear function of $s$ around $d\sim 0.4$, with the numerical solution of $\hat{r}$ denoted as $\hat{r}_n$ and the fitted solution denoted as $\hat{r}_f$, the result is $\hat{r}_n\approx \hat{r}_{f}$ with
\begin{equation}
\begin{split}
&\hat{r}_f\approx 1+\left(105.837 b^2-197.435 bd+55.6138 b+62.0744 d^2-22.544 d-0.873711\right) s\\
&+\left(-653.657 b^2+1108.47 bd-301.911 b-483.322 d^2+310.55 d-60.965\right)s^2\\
\end{split}
\label{eq.2.7}
\end{equation}
$\hat{r}_f$ is compared with $\hat{r}_n$ in Fig.~\ref{fig.fitrhat}.
It can be found that, for most cases $\hat{r}$ is smaller than $1$ and is decreasing with the growth of $s$, which indicates that the skyrmions become smaller with the growth of density even when $b$ and $d$ are unchanged.

We define the density of skyrmions as $\rho=N/A$ where $N$ is the number of skyrmions within an area $A$.
$R$ is approximately half of the average distance between the skyrmions, therefore $R$ can be related to $\rho$.
Assuming the skyrmions are distributed homogenously, then $N\approx A/\pi R^2$, and $s\approx \sqrt{\pi N / A}$.

It has been found that, by using the leading order ansatz $\theta _{\rm LO}$, to minimize $F$ yields $\omega = w_0 b^2/d^2$ where $w_0=0.768548$ is a constant~\cite{radius2}.
By using Eq~(\ref{eq.2.6}), the average radius $r_s$ can be written as
\begin{equation}
r_s(b,d,N)=\sqrt{\frac{2}{w_0}\log \left(\frac{\pi }{\cos^{-1}(h)}\right)}\times \frac{d}{b} \times \hat{r}(s=\sqrt{\frac{\pi N}{A}}).
\label{eq.4.2}
\end{equation}
In the following, we use $h=0.9$, $\sqrt{2\log \left(\pi /\cos^{-1}(h)\right)/w_0}\approx 2.24744$, using $\hat{r}_f$ to approximate $\hat{r}$,
\begin{equation}
\begin{split}
&r_s(b,d,\rho)\approx \frac{2.247d}{b}\left\{ 1+\left[d (975.6 -1518.4 d)-2053.5 b^2+b (3482.4 d-948.481)-191.5\right]\rho\right.\\
&\left.+\left[b (98.57 -349.9 d)+187.6 b^2+d (110.0 d-39.96)-1.549\right]\sqrt{\rho}\right\}.\\
\end{split}
\label{eq.4.3}
\end{equation}

In Ref.~\cite{radius4}, the equilibrium $R$ is numerically solved by minimize the energy density.
Note that $R$ in this case is independent of the density of the skyrmions.
In our case, $R$ is a quantity between the case of skyrmion lattice and the case of a single isolated skyrmion, and is determined by the density of the skyrmions, one can calculate $r_s$ after $R$ is given.

\section{\label{level3}Lattice simulation}

The lattice simulation is based on the LLG equation, denoting ${\bf n}_{\bf r}$ as the local magnetic momentum at site ${\bf r}$, the LLG can be written as~\cite{latticesimulation,otherlatticesimulation,llg1,llg2}
\begin{equation}
\frac{d}{dt} {\bf n}_{\bf r}=-{\bf B}_{\rm eff}({\bf r})\times {\bf n}_{\bf r}-\alpha {\bf n}_{\bf r}\times \frac{d}{dt} {\bf n}_{\bf r},\\
\label{eq.3.1}
\end{equation}
where ${\bf n}_{\bf r}$ is the local magnetic moment, $\alpha$ is the Gilbert damping constant and the effective magnetic field ${\bf B}_{\rm eff}$ is
\begin{equation}
{\bf B}_{\rm eff}({\bf r})=-\frac{\delta H}{\delta {\bf n}_{\bf r}},
\label{eq.3.2}
\end{equation}
with the discretized version of Hamiltonian defined as~\cite{hamitonian2,jdb1}
\begin{equation}
H=\sum _{{\bf r},i=x,y}\left[-J({\bf r}){\bf n}_{{\bf r}+\delta_i}-D({\bf r}) {\bf n}_{{\bf r}+\delta_i}\times {{\bf e}}_i-{\bf B}\right]\cdot {\bf n}_{\bf r},\\
\label{eq.3.3}
\end{equation}
where $\delta _{i}$ refers to each neighbour. On a square lattice, one has $\delta _{i}= {\bf e}_{i}$, therefore
\begin{equation}
{\bf B}_{\rm eff}({\bf r})=\sum _{i=x,y}\left[J({\bf r}){\bf n}_{{\bf r}+\delta_i}+J({\bf r}-\delta_i){\bf n}_{{\bf r}-\delta_i}\right]
+\sum _{i=x,y}\left[D({\bf r}){\bf n}_{{\bf r}+\delta_i}\times {{\bf e}}_i-D({\bf r}-\delta _i){\bf n}_{{\bf r}-\delta_i}\times {{\bf e}}_i\right]+{\bf B}({\bf r}).
\label{eq.3.4}
\end{equation}
The simulation was carried out on GPU~\cite{latticesimulation} which has a great advantage over
CPU because of the ability of parallel computing of the GPU. Eq.~(\ref{eq.2.1}) is numerically integrated by using the fourth-order Runge-Kutta method.

\section{\label{level4}Numerical results}

We run the simulation on a $512\times 512$ square lattice.
In the simulation, we use dimensionless homogeneous $J$, $D$ and $B$.
$J=1$ is used as the definition of the energy unit~\cite{jdb1,jdb2,jdb3}, the results are presented with $d$ and $b$.
In the previous works, the Gilbert constant was chose to be $\alpha=0.01$ to $1$~\cite{otherlatticesimulation,llg2,jdb2,jdb3,alpha1,alpha2,alpha3,alpha4,alpha5,alpha6,alpha7,alpha8}.
In this work, we use $\alpha=0.04$ which is in the region of commonly used $\alpha$.
The time step is denoted as $\Delta t$. We use $\Delta t=0.01$ time unit, and the configurations are stabled typically after about $10^6\sim 10^7$ steps starting with a randomized initial state.
The average radius of the skyrmions is measured as $r_s=A_s/NA$ where $A_s$ is the total area of the skyrmions which is determined by the number of sites in the isoheight $n_z=h$ with $h=0.9$, $A=512^2$ and $N$ is the number of skyrmions.
The standard error of the radii of skyrmions are also measured.

\begin{figure}[!htbp]
\centering
\includegraphics[width=0.7\textwidth]{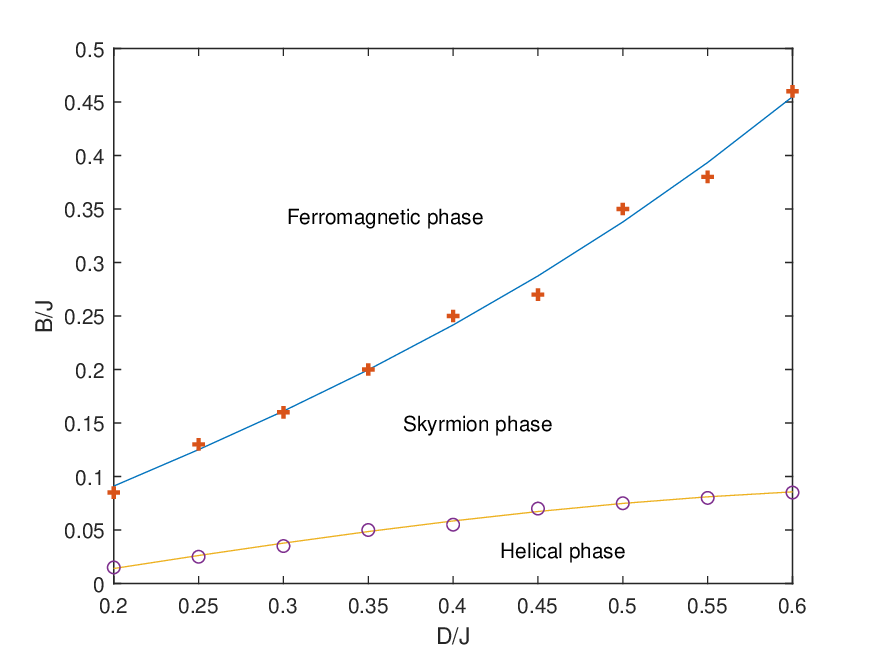}
\caption{\label{fig.phasediagram}The phase diagram obtained by lattice simulation of LLG with randomized initial states.}
\end{figure}

To investigate the relationship between $r$, $d$ and $b$, we simulate with $d$ in the range of $0.2$ to $0.6$, and with growing $b$ for each fixed $d$.
We focus on those configurations that are in the skyrmion phase when stable.
The phase diagram is shown in Fig.~\ref{fig.phasediagram}.

\subsection{\label{level4.1}The relationship between the average radius and density}

\begin{figure}[!htbp] \centering
\subfigure[$N=1133$]{\includegraphics[width=0.32\textwidth]{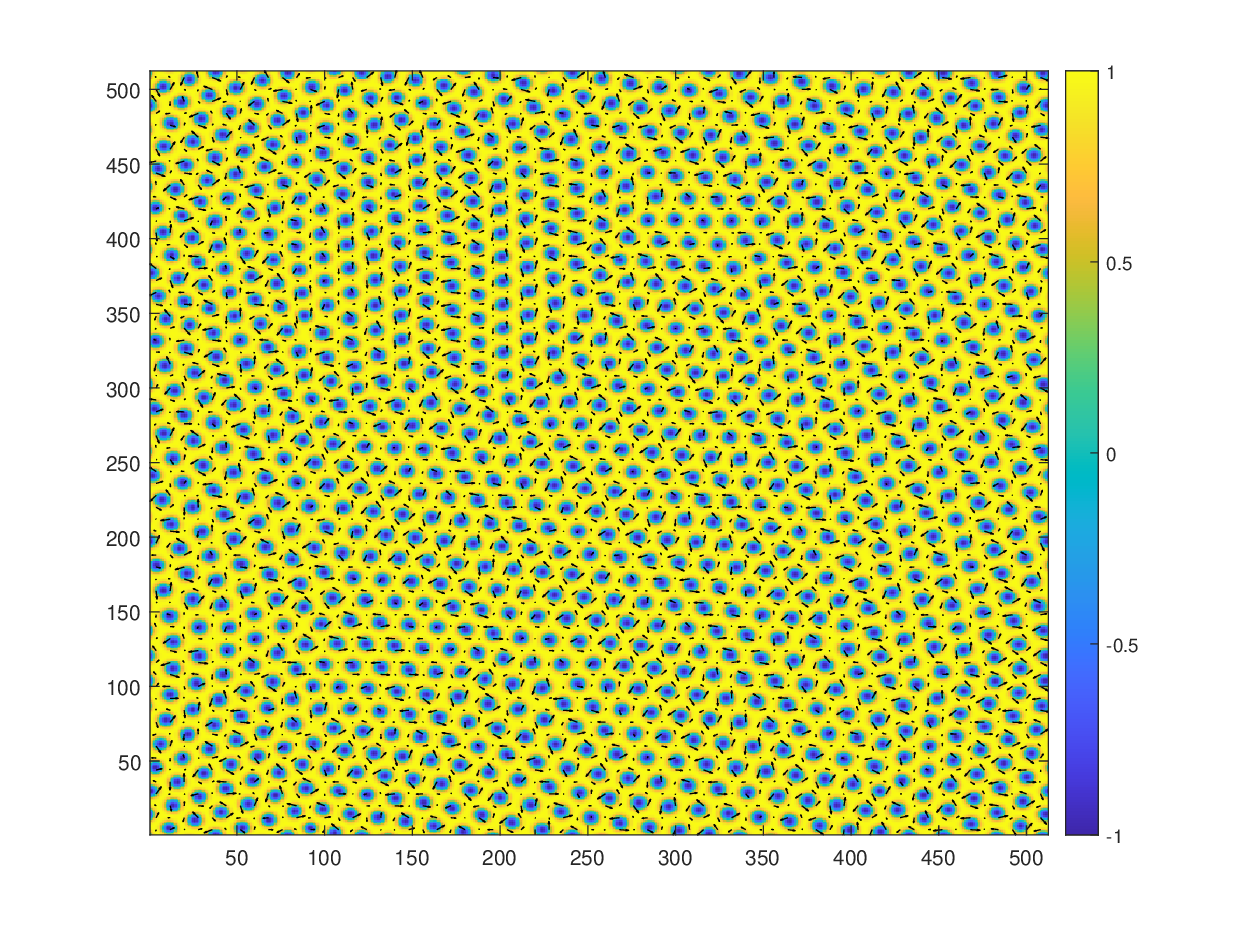}}
\subfigure[$N=1026$]{\includegraphics[width=0.32\textwidth]{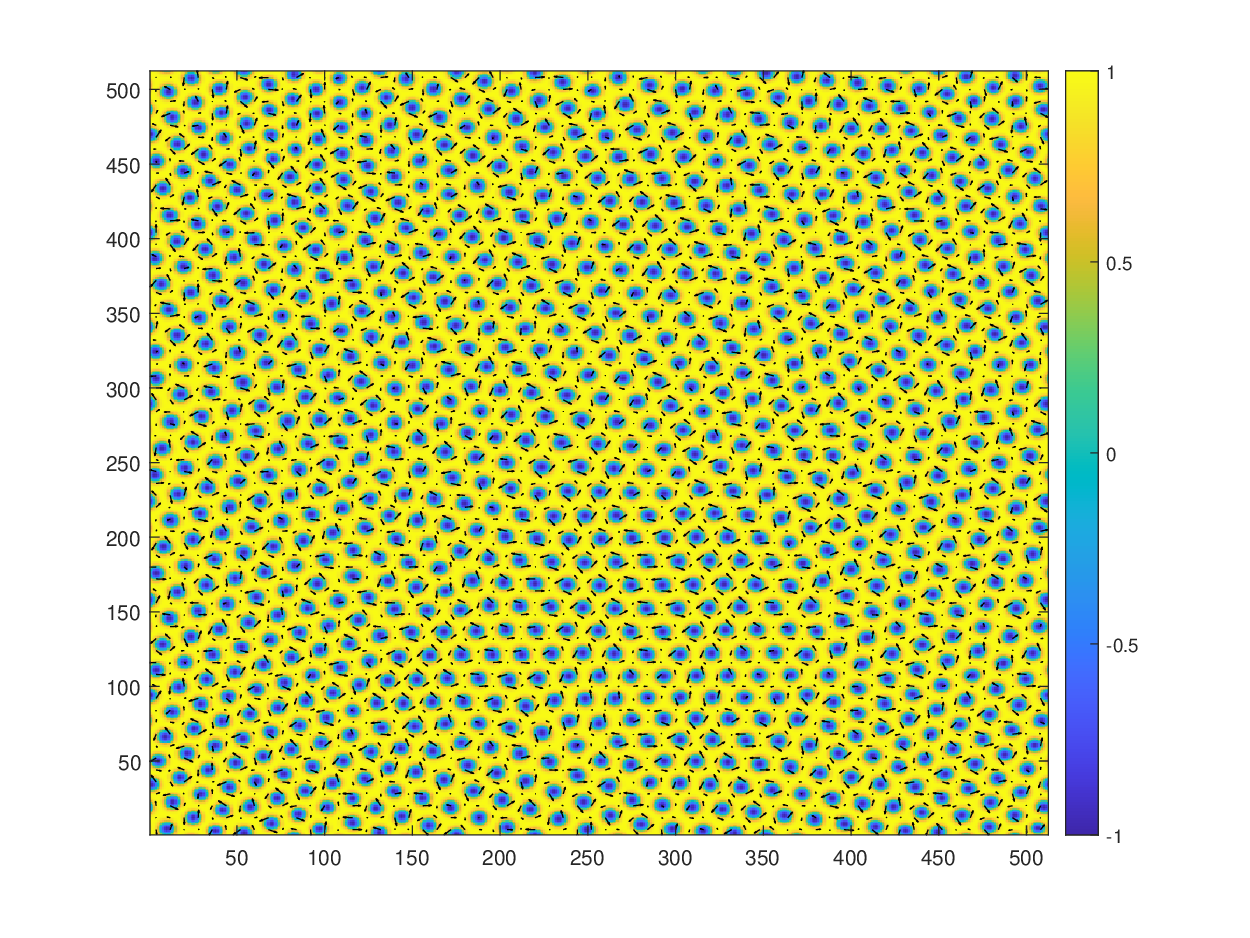}}
\subfigure[$N=919$]{\includegraphics[width=0.32\textwidth]{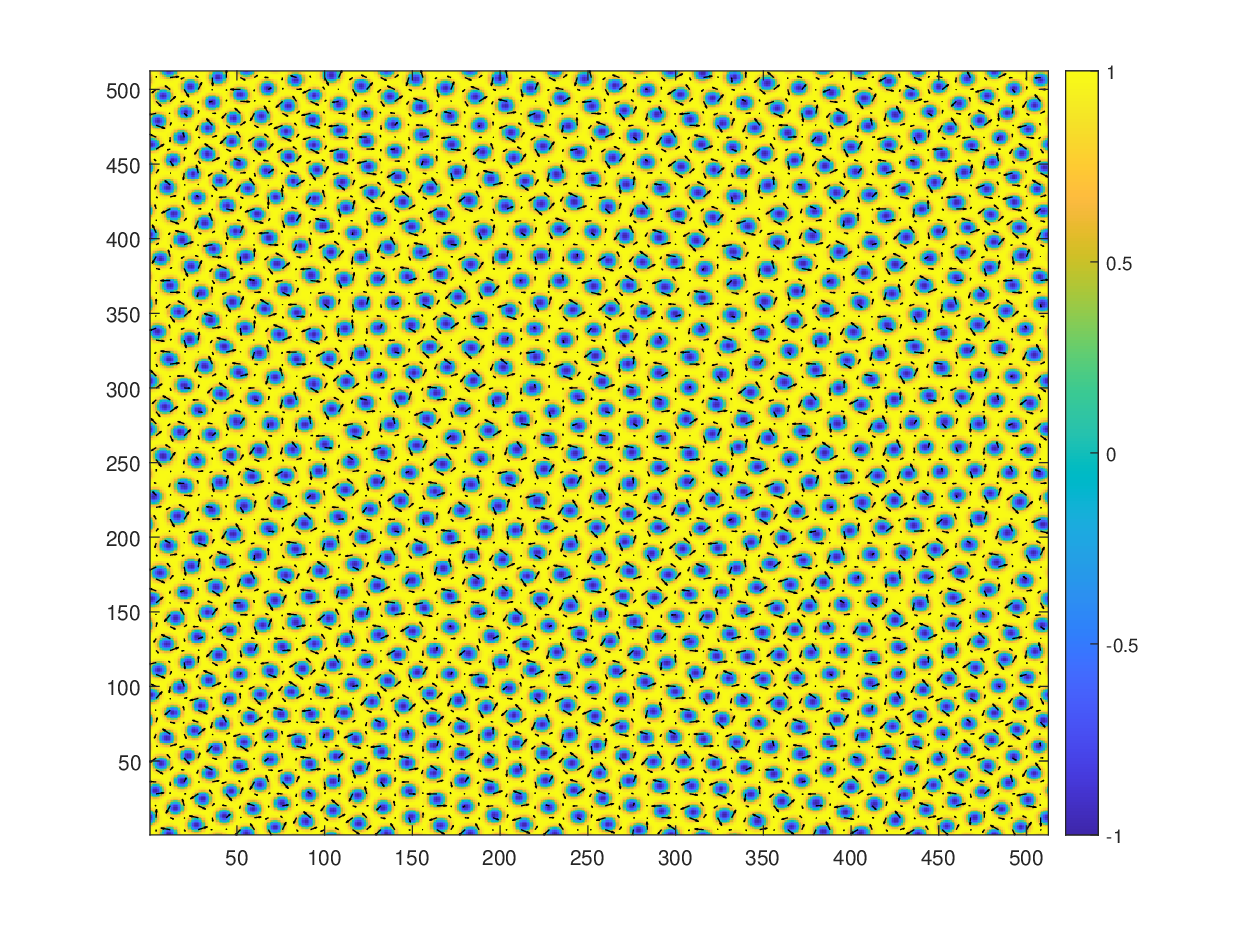}}
\subfigure[$N=787$]{\includegraphics[width=0.32\textwidth]{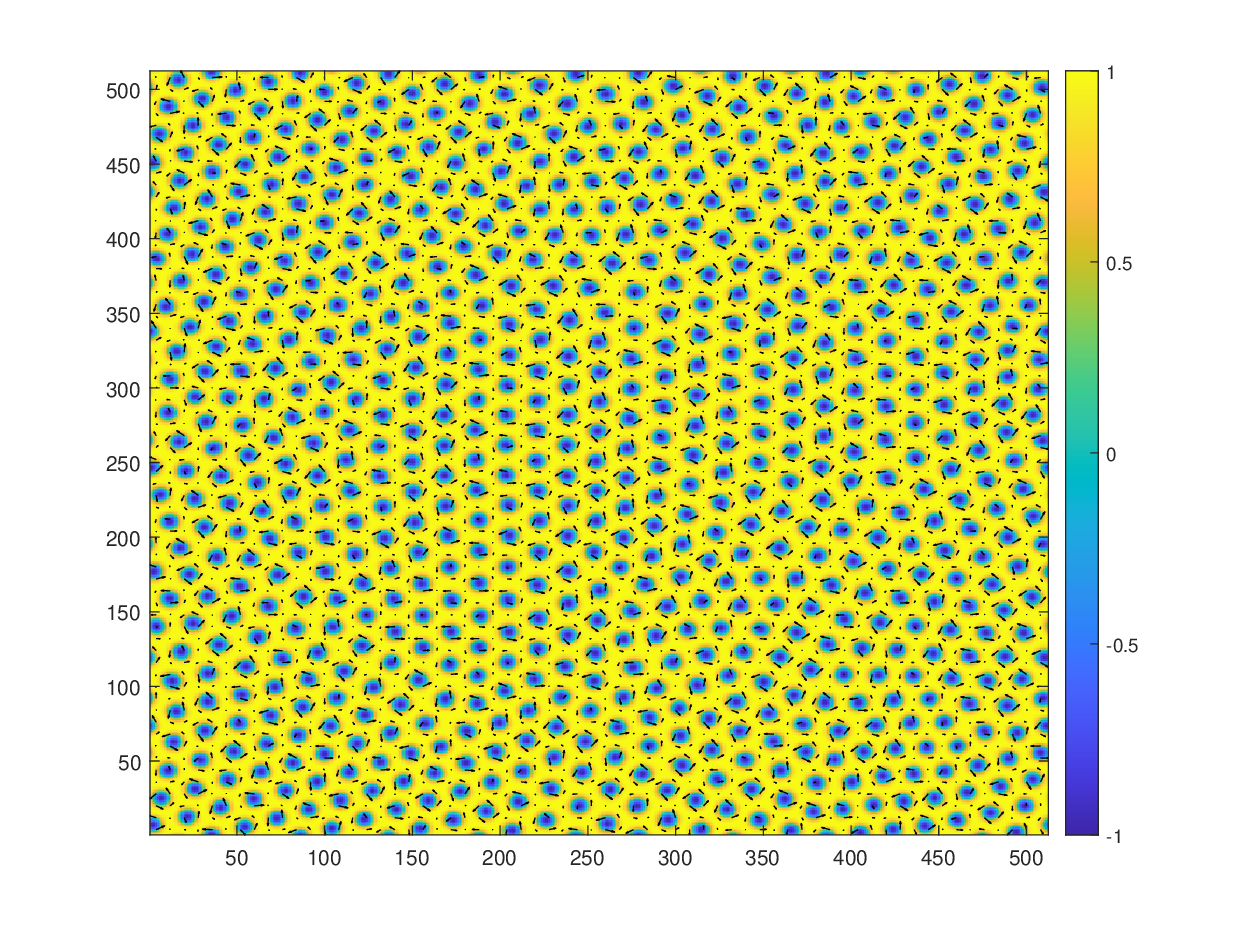}}
\subfigure[$N=646$]{\includegraphics[width=0.32\textwidth]{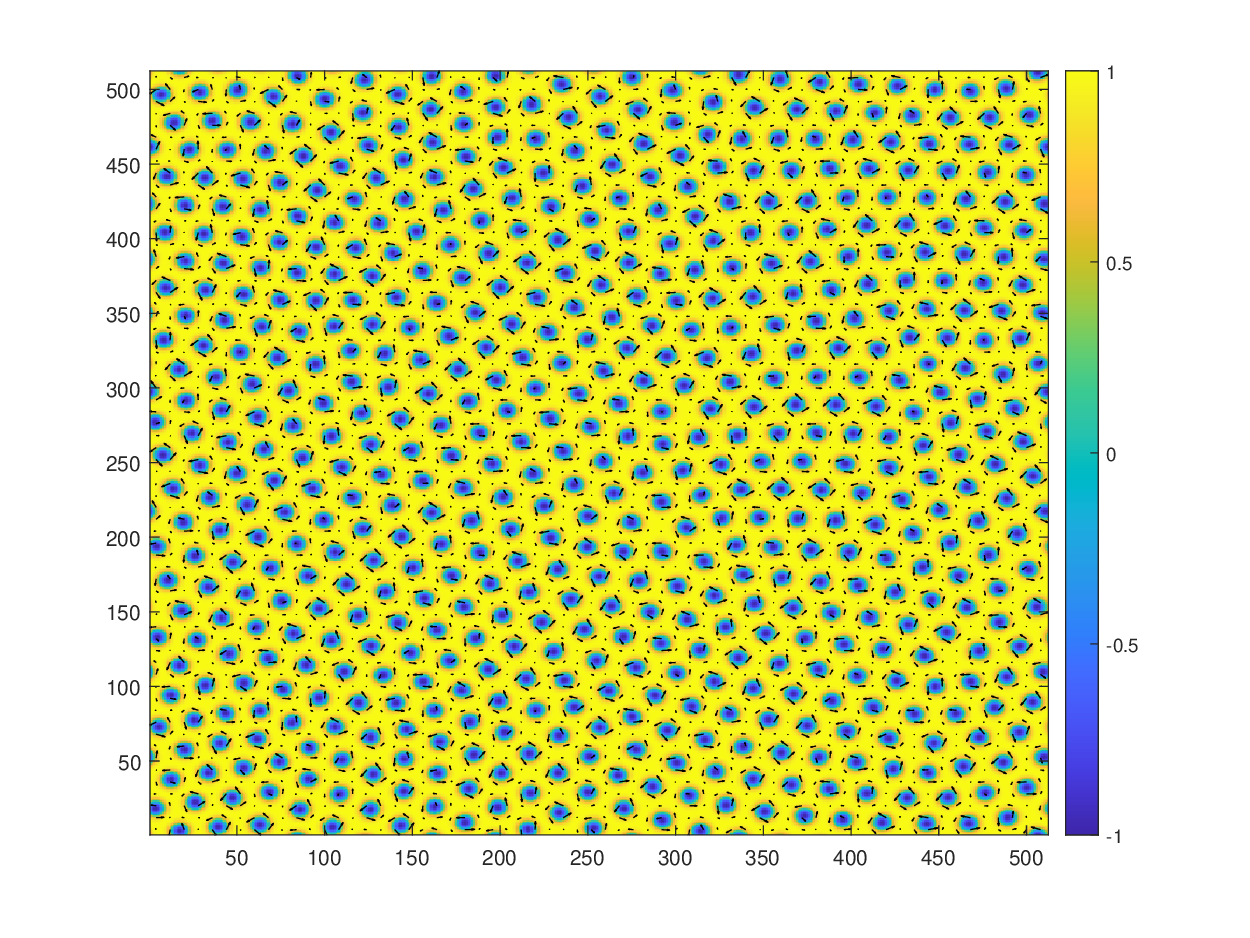}}
\subfigure[$N=595$]{\includegraphics[width=0.32\textwidth]{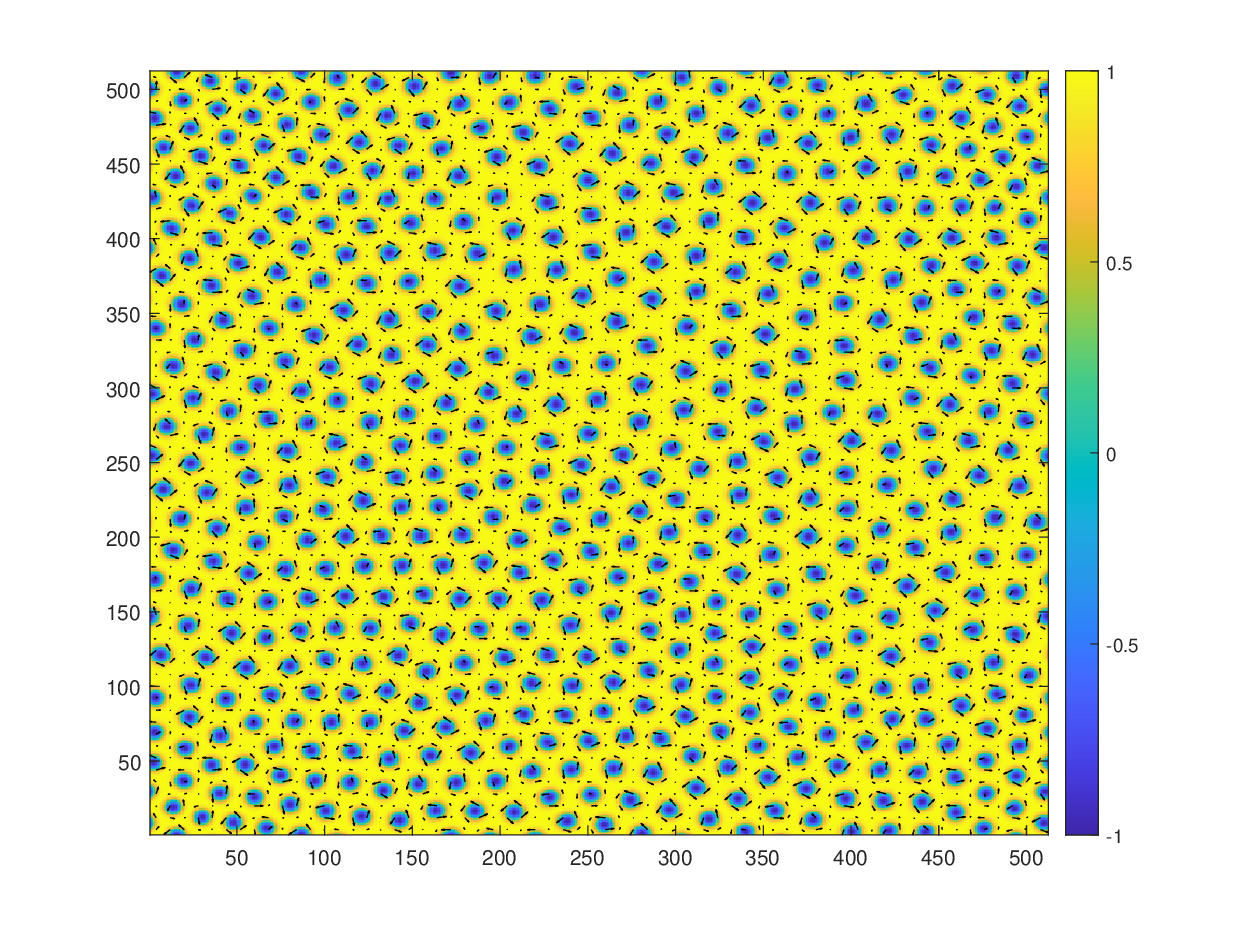}}
\subfigure[$N=457$]{\includegraphics[width=0.32\textwidth]{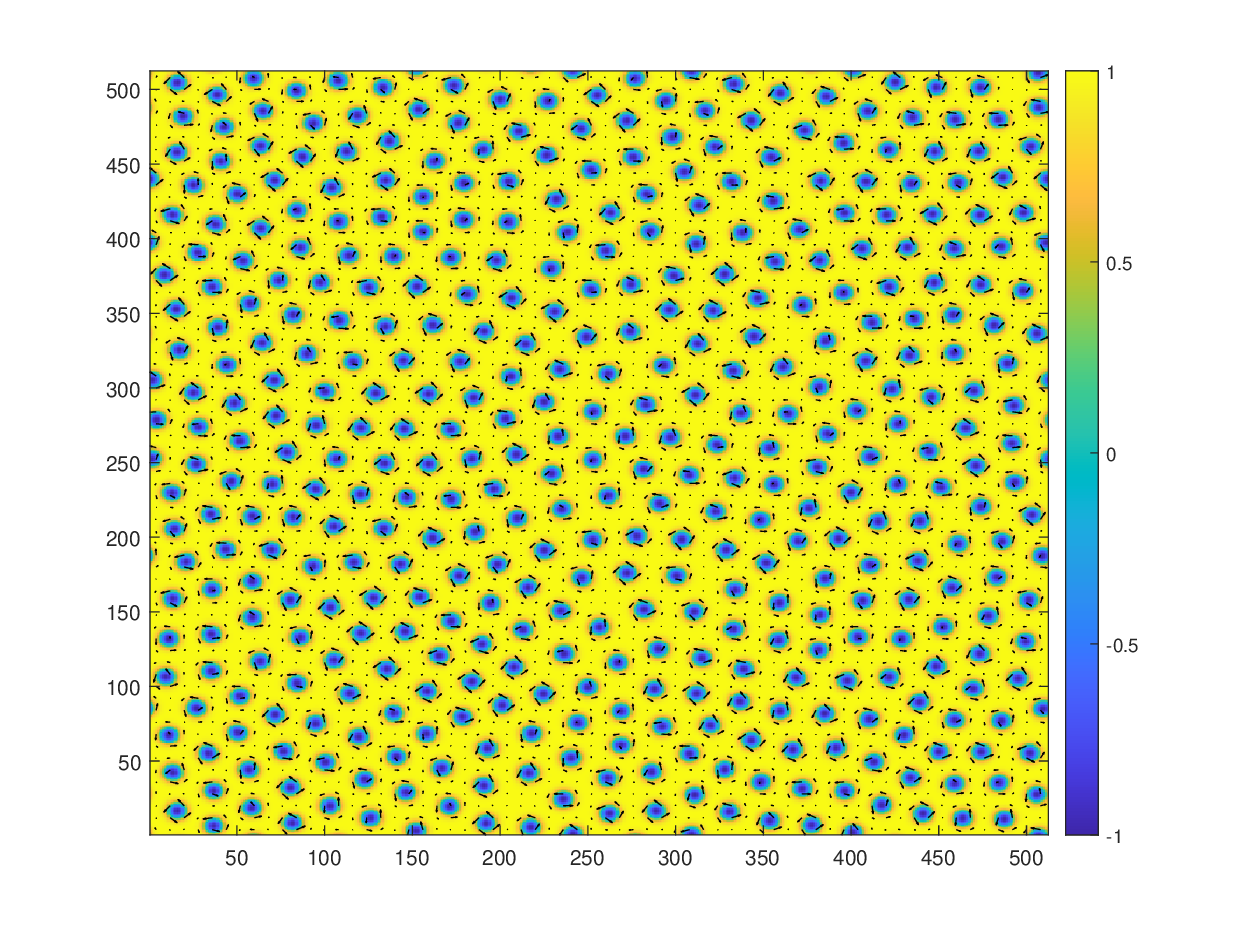}}
\subfigure[$N=358$]{\includegraphics[width=0.32\textwidth]{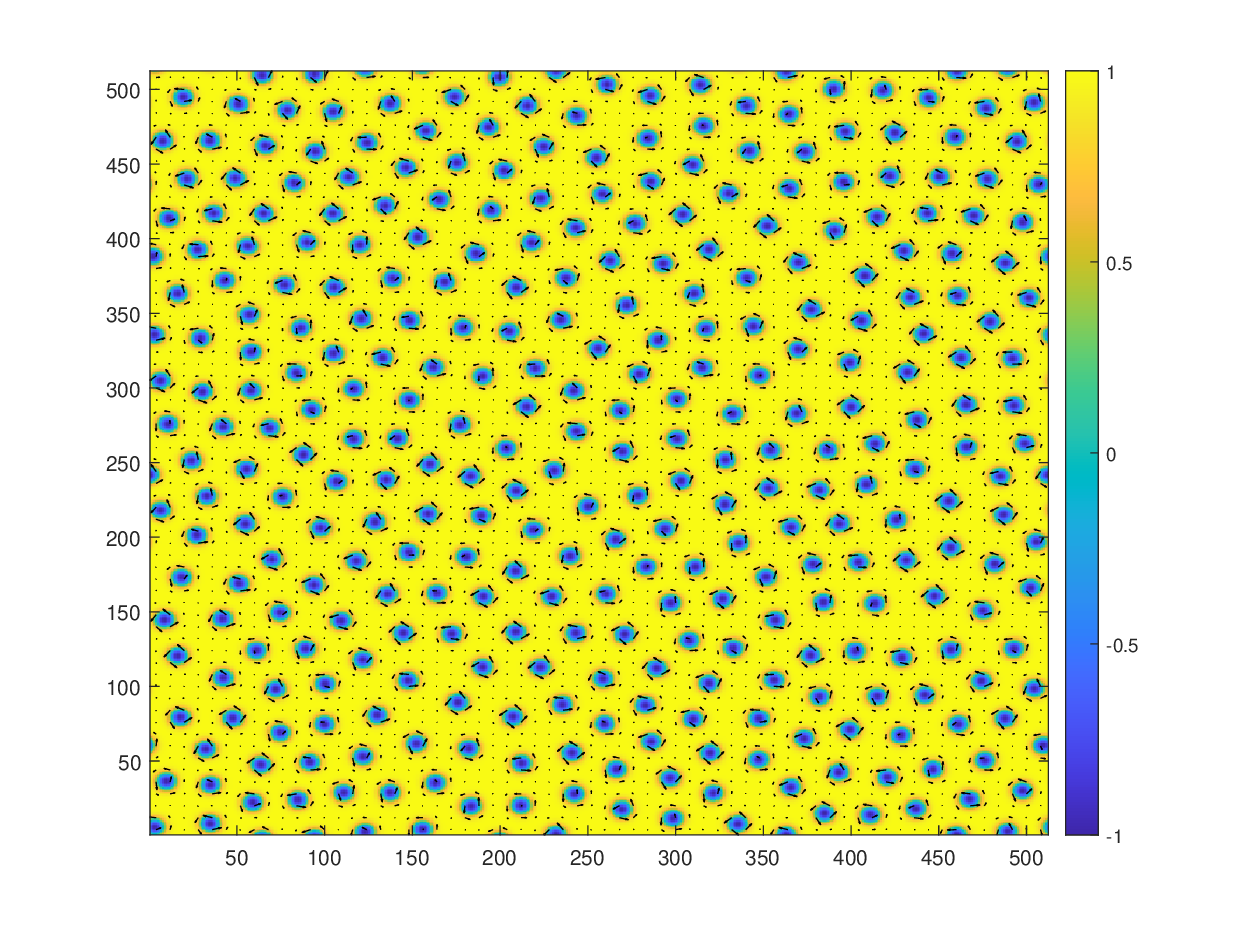}}
\subfigure[$N=238$]{\includegraphics[width=0.32\textwidth]{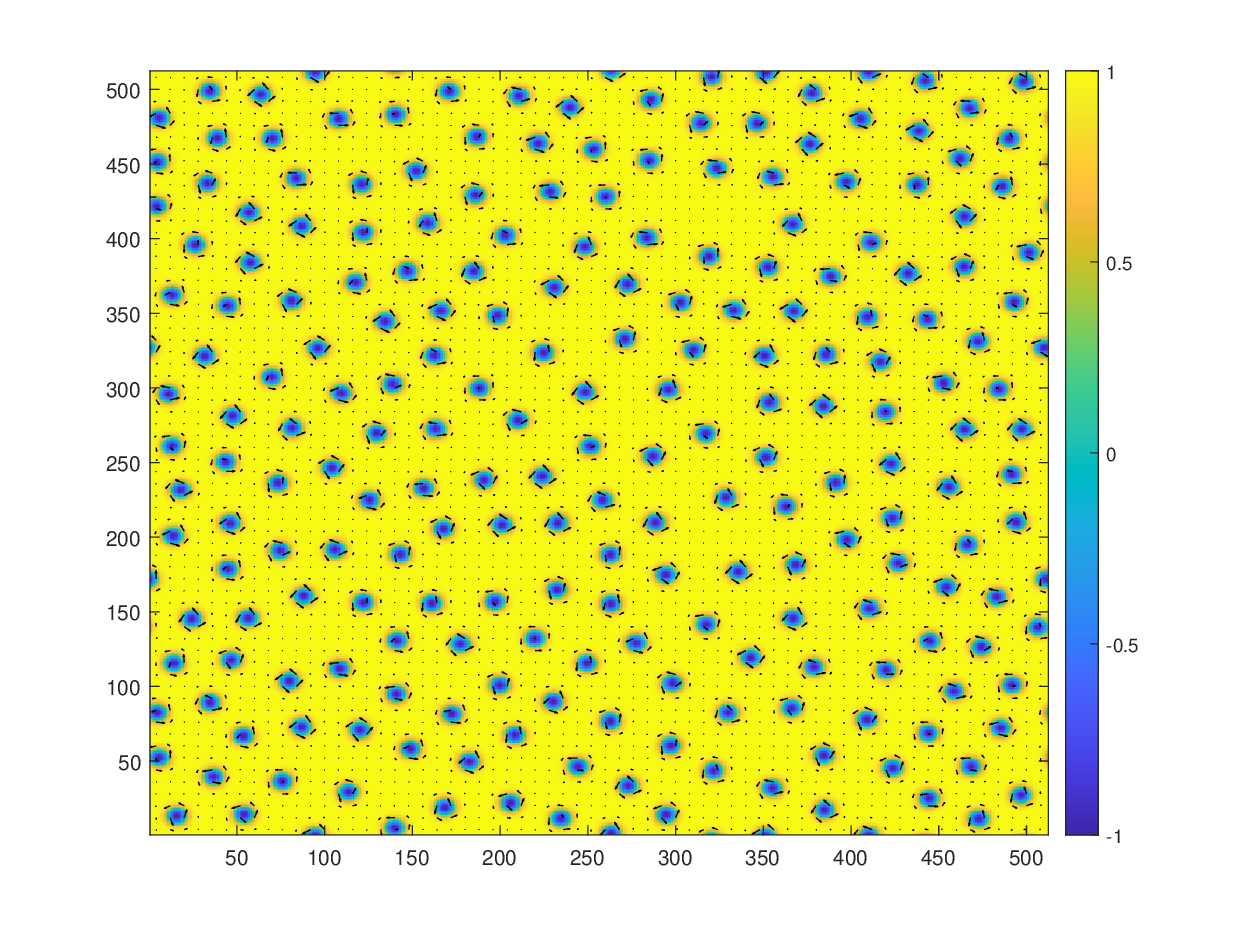}}
\subfigure[$N=123$]{\includegraphics[width=0.32\textwidth]{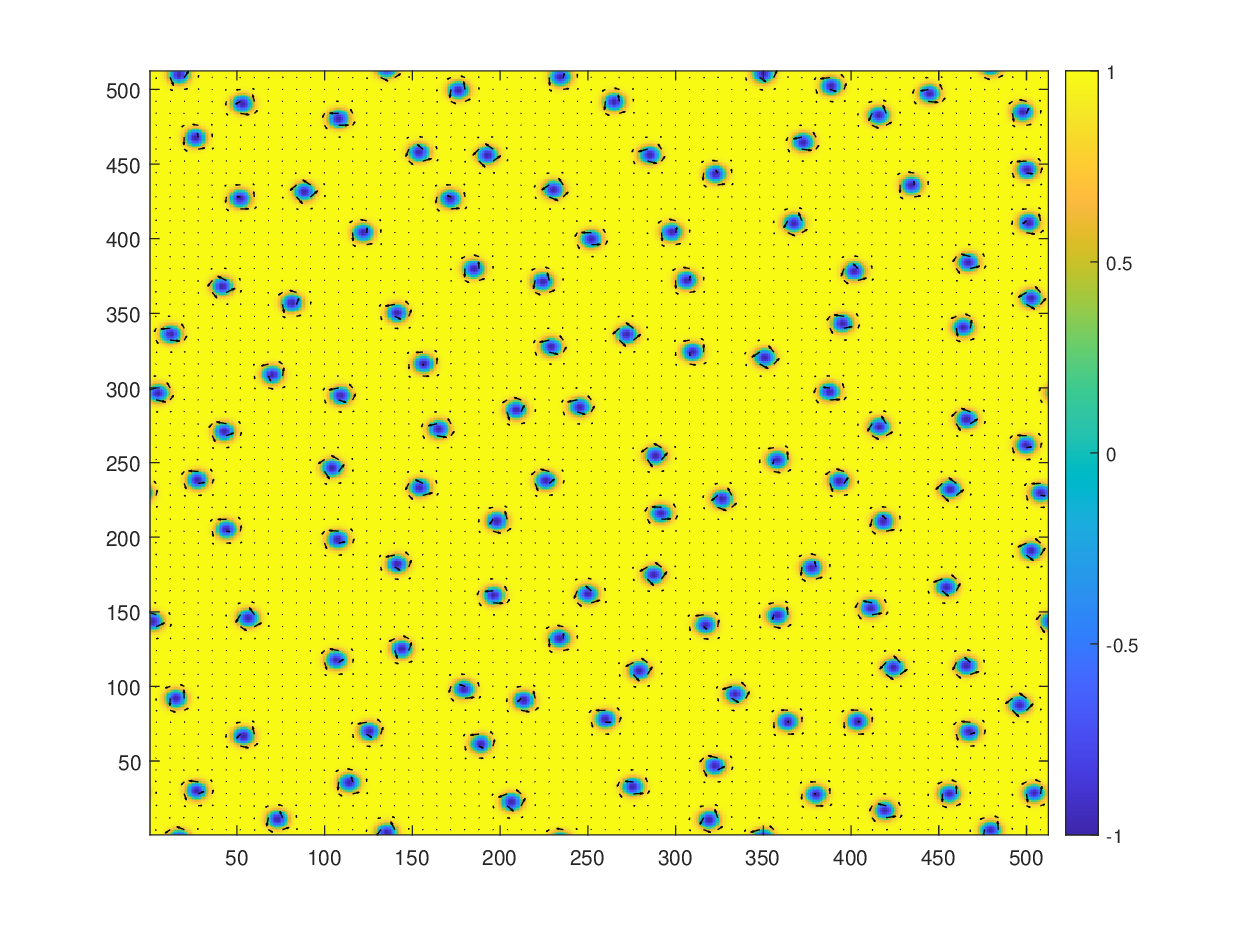}}
\subfigure[$N=1$]{\includegraphics[width=0.32\textwidth]{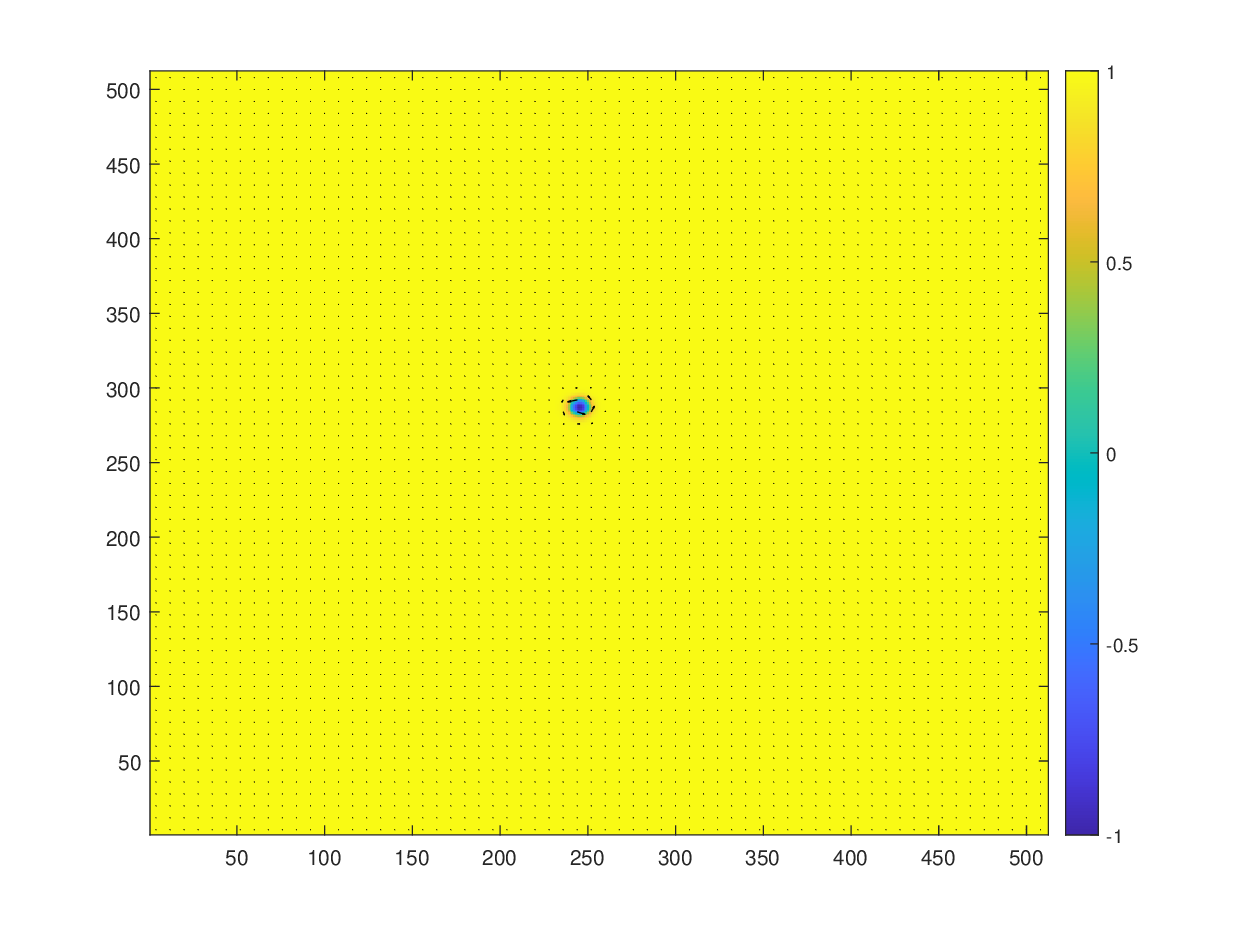}}
\caption{{\label{conf}The configurations corresponding to different $N$.}}
\end{figure}

In this subsection, we use the configurations at $d=0.2,b=0.025$, $d=0.35,b=0.1$, $d=0.4,b=0.1$, $d=0.4,b=0.2$, $d=0.45,b=0.15$ and $d=0.6,b=0.15$ to investigate the relationship between the average radius and density.
Firstly we calculate the number of skyrmions in each configuration.
By repeatedly and randomly erasing about $10\%$ of the total number of skyrmions at a time and performing the simulation sequentially, we obtain the configurations at different $N$.
Taking the case of $d=0.4, b=0.1$ as an example, the resulting configurations are shown in Fig.~\ref{conf}.

\begin{figure}[!htbp] \centering
\subfigure[$d=0.45,b=0.15$]{\includegraphics[width=0.48\textwidth]{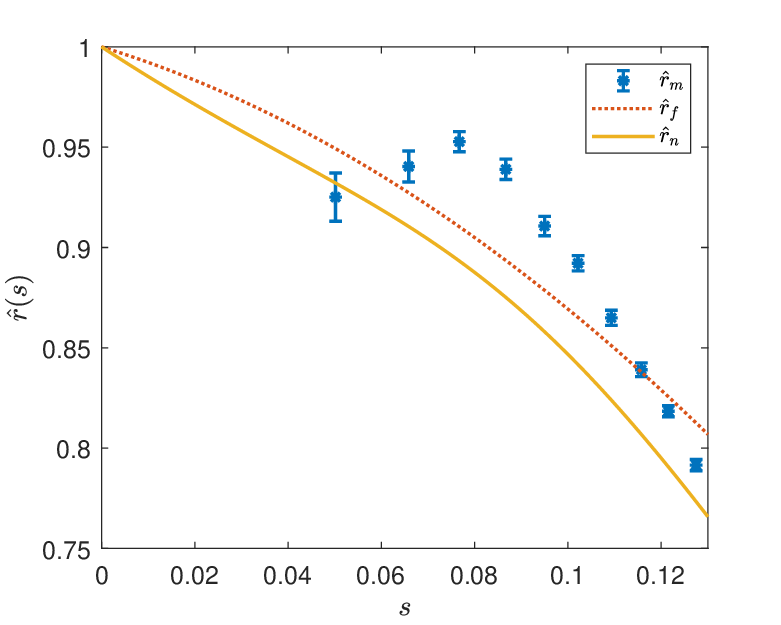}}
\subfigure[$d=0.35,b=0.1$]{\includegraphics[width=0.48\textwidth]{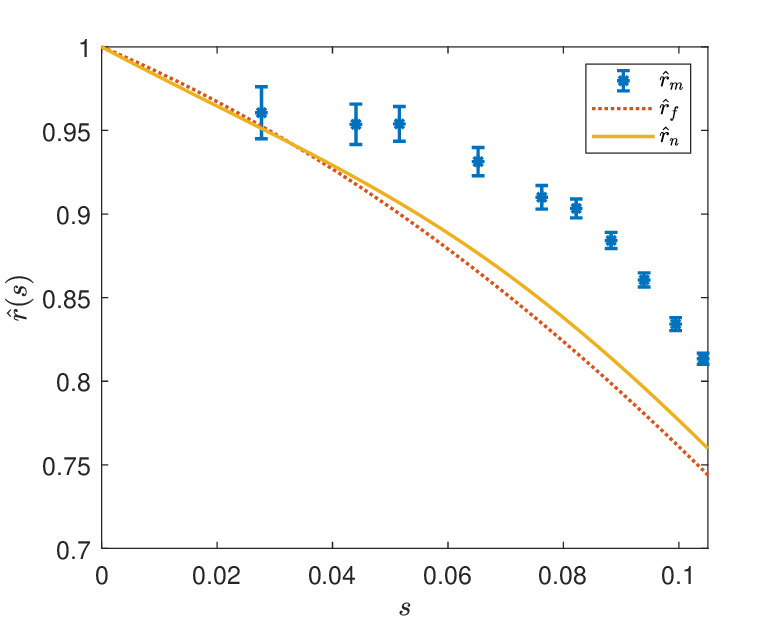}}\\
\subfigure[$d=0.4,b=0.1$]{\includegraphics[width=0.48\textwidth]{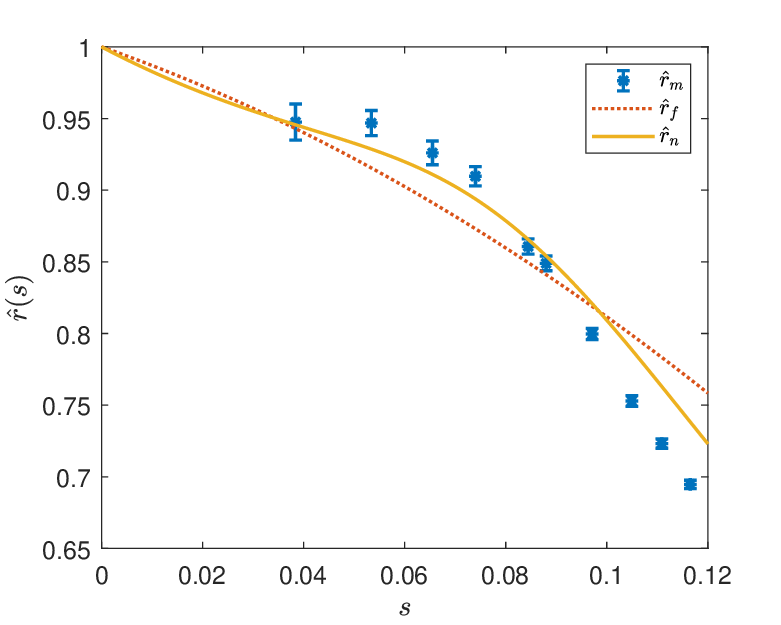}}
\subfigure[$d=0.2,b=0.025$]{\includegraphics[width=0.48\textwidth]{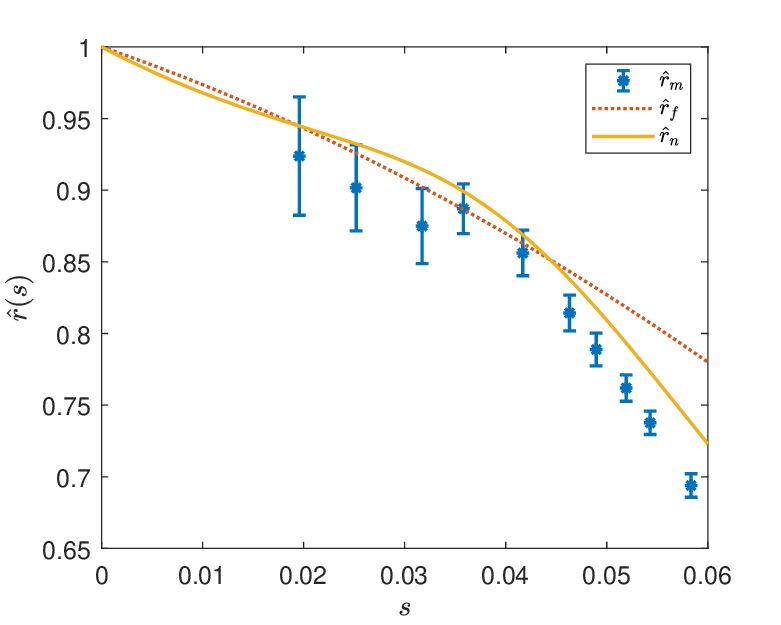}}
\caption{{\label{fitwell}Compare $\hat{r}_m$ with $\hat{r}_n$ and $\hat{r}_f$.}}
\end{figure}

Because the size of the lattice is $512\times 512$, $s\approx \sqrt{N\pi/512^2}\approx 0.00346 \sqrt{N}$.
The ratios of the average radii of skyrmions to the radius of an isolated skyrmion at different $N$ are measured and denoted as $\hat{r}_m$.
We compare $\hat{r}_m$, $\hat{r}_f$ and $\hat{r}_n$ in Fig.~\ref{fitwell}.
It can be seen that the theocratical results $\hat{r}_n$ and $\hat{r}_f$ can approximately predict $\hat{r}_m$ correctly, the deviations between $\hat{r}_m$ and $\hat{r}_{s,f}$ are generally within about $10\%$.
It can be seen that, generally, for larger skyrmions, the theocratical results are better.
Especially, for $d=0.4, b=0.1$, $\hat{r}_s$ can fit $\hat{r}_m$ very well.
For the cases where the sizes of skyrmions are relatively smaller, there are several possible reasons for the deviation.
On one hand, when the skyrmions are smaller, they are not homogenously aligned, the distances between the skyrmions become larger, consequently the actual $s$ is smaller for $\hat{r}_m$, so the points of $\hat{r}_m$ are biased towards a larger $s$.
Secondly, the lattice simulation is more coarse for smaller skyrmions, which can also lead to differences with the theoretical results.
Similarly, if the skyrmions are small and occupy only hundreds of sites, the $\theta(r)$ can no longer be treated as a continuous function.

\begin{figure}[!htbp] \centering
\includegraphics[width=0.48\textwidth]{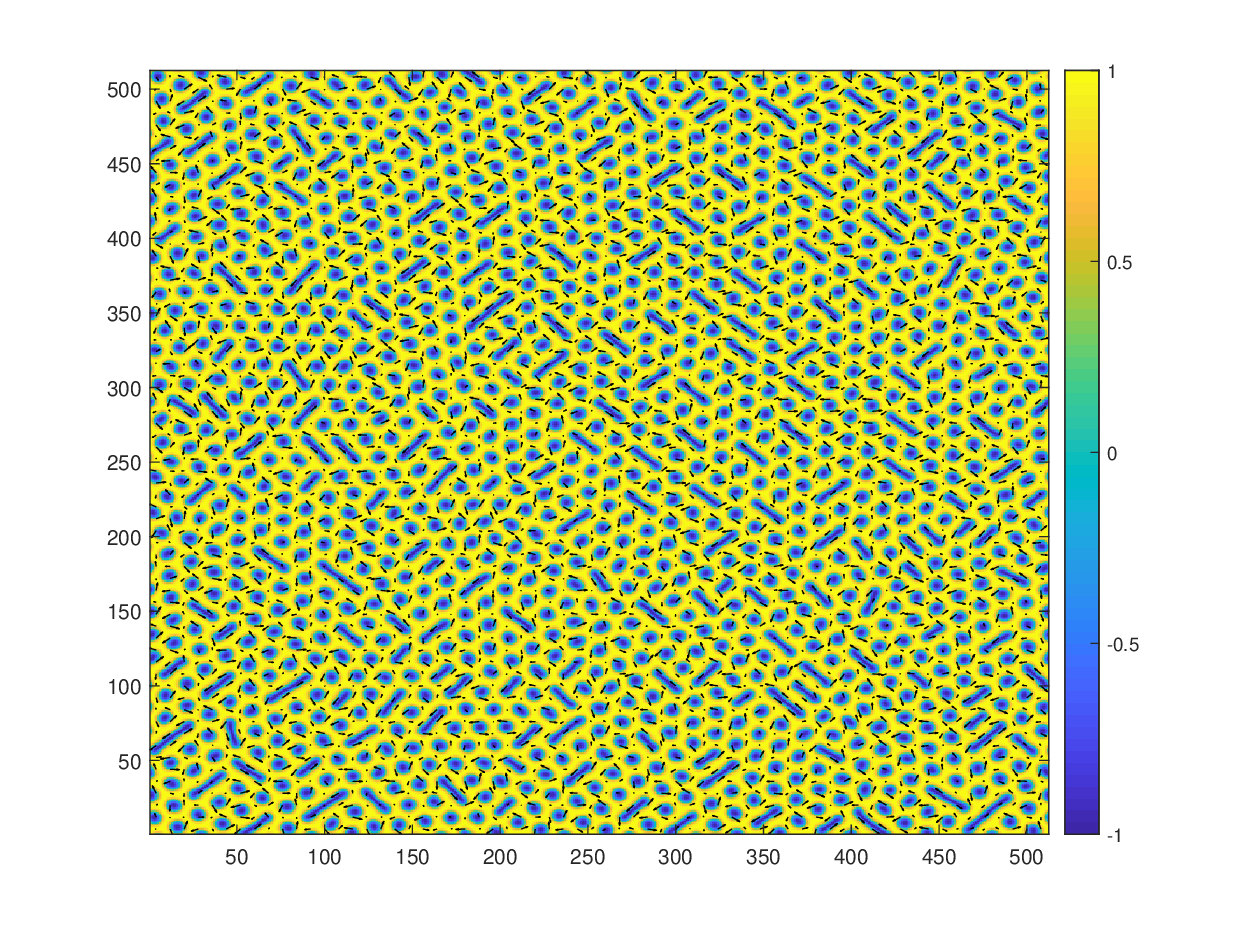}
\includegraphics[width=0.48\textwidth]{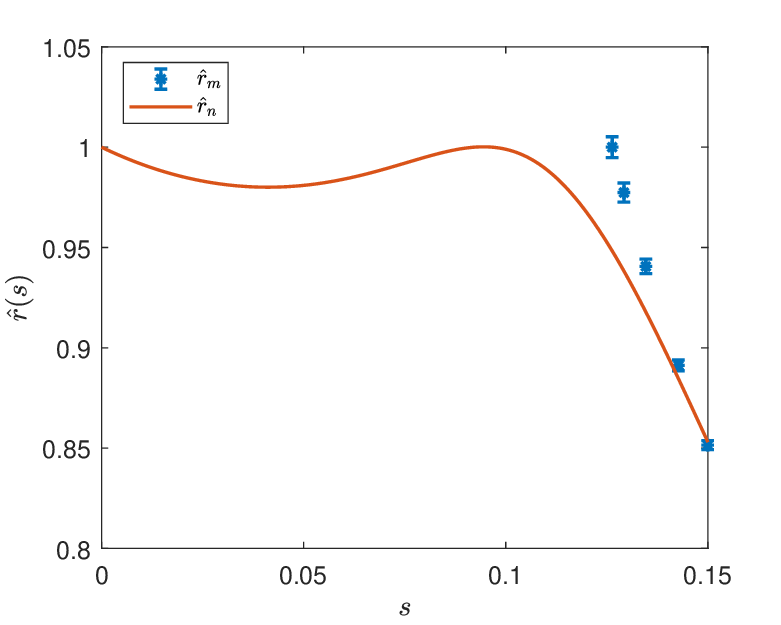}
\caption{{\label{fitnotwell06015}Compare $\hat{r}_m$ with $\hat{r}_n$ for $d=0.6, b=0.15$.}}
\end{figure}

There are also cases that the $\hat{r}_m$ are not fitted very well by the theocratical predictions.
In the case of $d=0.6, b=0.15$, after erasing some skyrmions, the configuration began to enter the helical phase as shown in Fig.~\ref{fitnotwell06015}.~(a).
In this case, when an isolated skyrmion is created, it is not stable and will grow into stripes.
If we choose the average radius near the phase transition as a baseline, that is, we choose the $\hat{r}_m$ near the phase transition as $1$, the results are shown in Fig.~\ref{fitnotwell06015}.~(b).
It can be seen that the $\hat{r}_m$ approaches $\hat{r}_n$.

\begin{figure}[!htbp] \centering
\includegraphics[width=0.48\textwidth]{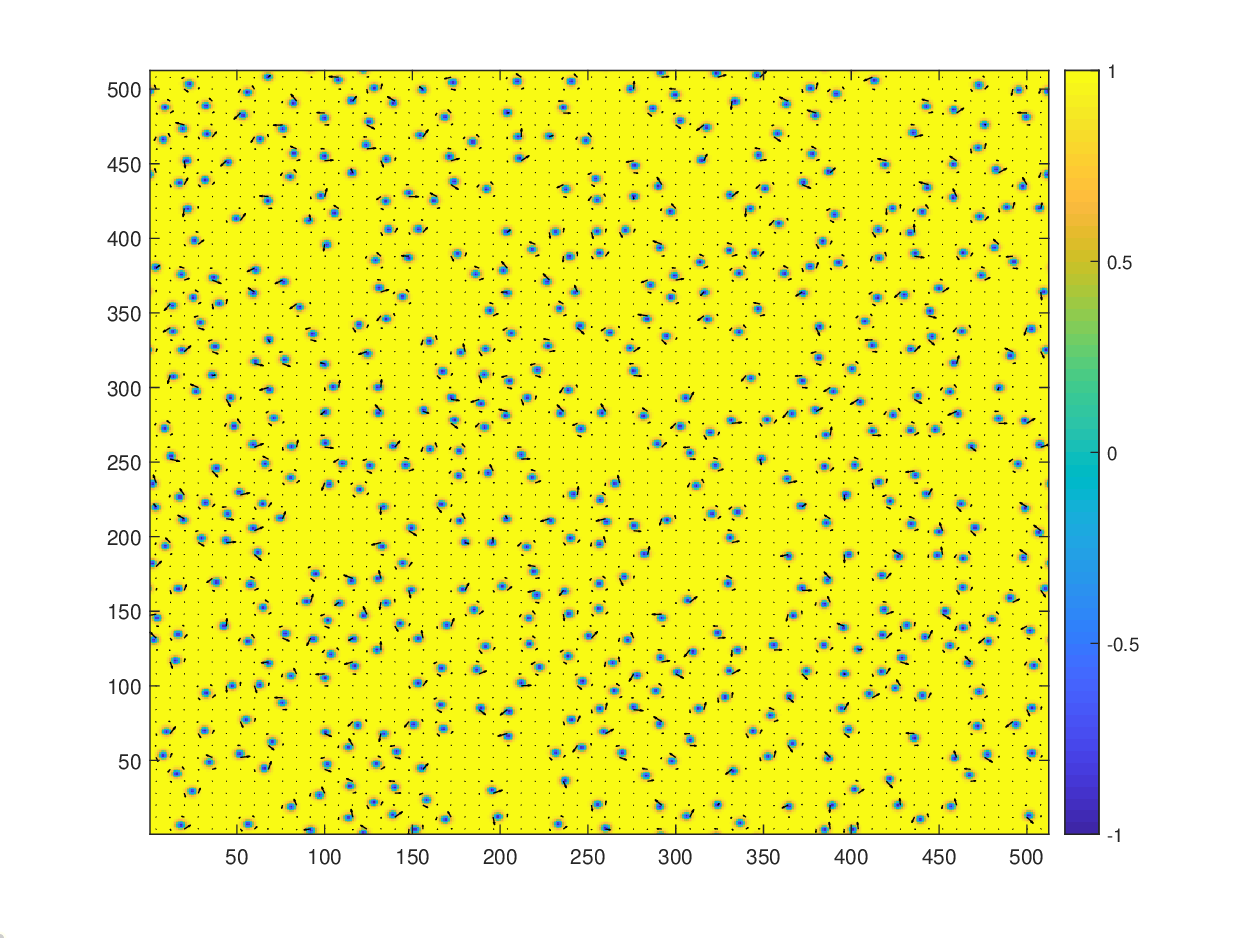}
\includegraphics[width=0.48\textwidth]{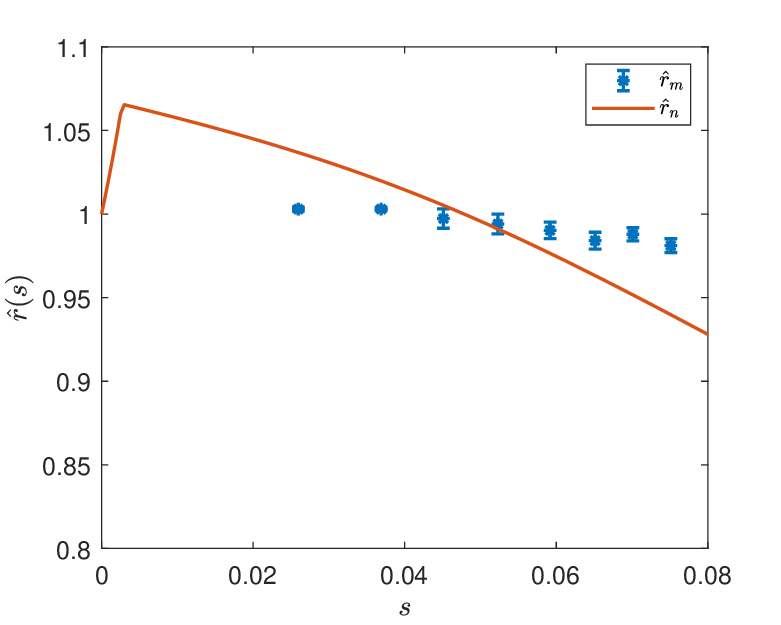}
\caption{{\label{fitnotwell0402}Compare $\hat{r}_m$ with $\hat{r}_n$ for $d=0.4, b=0.2$.}}
\end{figure}

Another case is when $d=0.4, b=0.2$ as shown in Fig.~\ref{fitnotwell0402}.
This is an example that when the skyrmions are small, they are no longer homogenously aligned.
It can been found from Fig.~\ref{fig.phasediagram} that the configuration is also near the phase transition between the skyrmion phase to the ferromagnetic phase.
It is interesting that, the average radius is not always decreasing with $s$ especially when $s$ is small.
For $N=51$, $r_s=4.7003\pm 0.0054$, for $N=103$, $r_s=4.7005\pm 0.0043$, which are larger than the case of a signle isolated skyrmion $r_s=4.6865$.
One can see that $\hat{r}_n$ has a similar behaviour.

\subsection{\label{level4.2}A formula for the average radius in the skyrmion phase}

\begin{figure}[!htbp]
\centering
\includegraphics[width=0.7\textwidth]{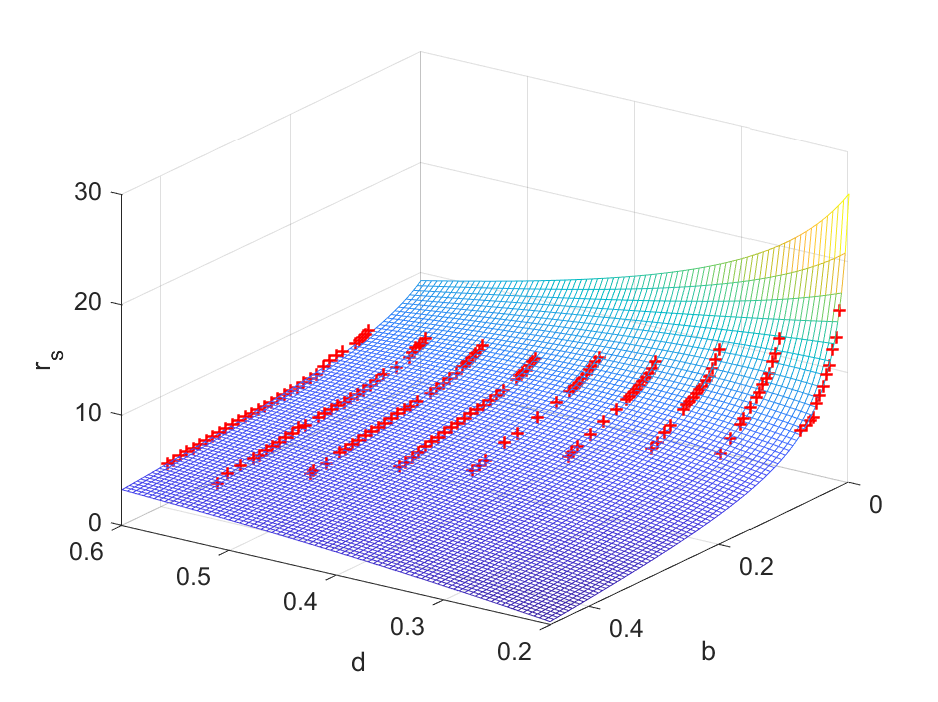}
\caption{\label{fig.fitres}$r_s$ at different $d$ and $b$ (marked as `+') and the fitted $r_s(d,b)$, i.e. Eq.~(\ref{eq.3.1}).}
\end{figure}
Since we start from the randomized initial state, the obtained configurations are the skyrmion lattices.
In this case, the relationship between $r_s$, $d$ and $b$ are fitted by a rational function, the result is
\begin{equation}
r_s(d,b)=\frac{-130.88 b^2+203.89 b d-25.2503 b-66.6575 d^2+65.0494 d+0.577576}{18.3341 b+3.88567 d-0.353591}.
\label{eq.4.1}
\end{equation}
$r_s$ and fitted $r_s(d,b)$~(Eq.~(\ref{eq.3.1})) are shown in Fig.~\ref{fig.fitres}.
One can see that the rational function is consistent with the numerical results.

\begin{figure}[!htbp]
  \centering
  \includegraphics[width=0.7\textwidth]{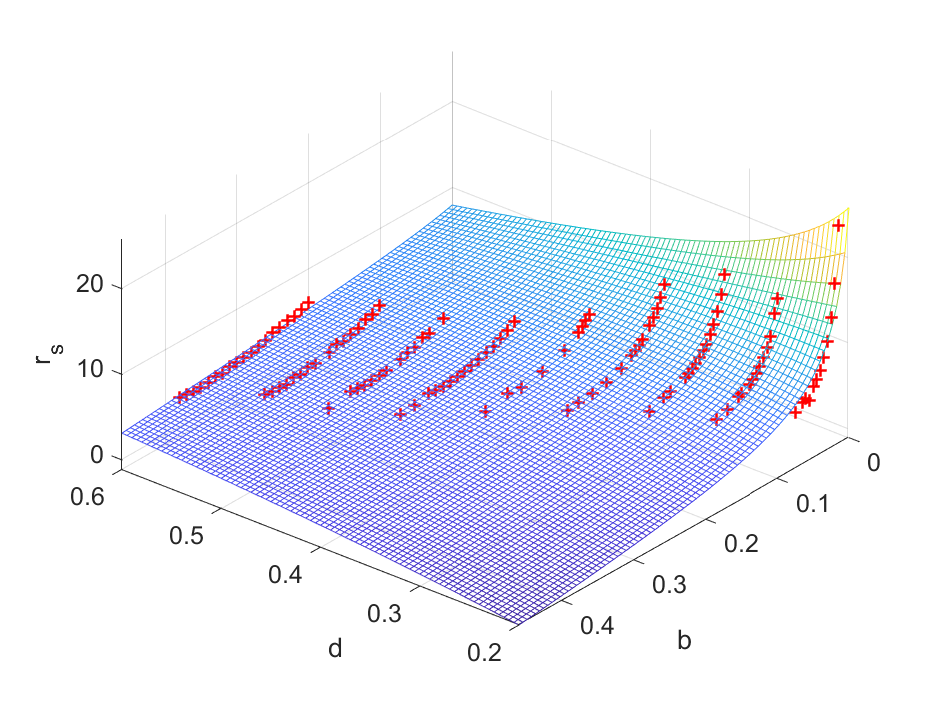}\\
  \caption{\label{correct2.fig}$r_s$ calculated with Eq.~(\ref{eq.4.3}) (marked as `+') compared with Eq.~(\ref{eq.4.1}) (the curved surface).}
\end{figure}

We compare Eq.~(\ref{eq.4.3}) with Eq.~(\ref{eq.4.1}) in Fig.~\ref{correct2.fig}.
Since the $\rho$ for each configuration is different, the results of $r_s(b,d,\rho)$ are depicted as points in Fig.~\ref{correct2.fig}.
Note that we remove the points correspond to the cases that the single isolated skyrmions are not stable such as $d=0.6$, $b=0.15$.
It can be seen that $r_s(b,d,\rho)$ can match the results very well.
$r_s(d,b,\rho)$ can be used to predict the average radius of the skyrmions when $d$, $b$ and the density is given.

\subsection{\label{level4.3}The shape of the skyrmion in the skyrmion phase}

\begin{figure}[!htbp]
  \centering
\subfigure[$d=0.45,b=0.15$]{\includegraphics[width=0.48\textwidth]{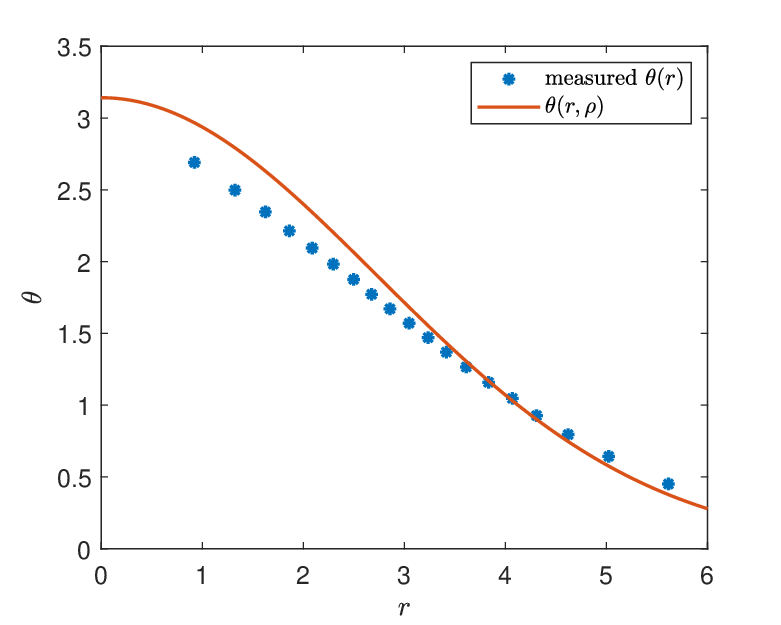}}
\subfigure[$d=0.35,b=0.1$]{\includegraphics[width=0.48\textwidth]{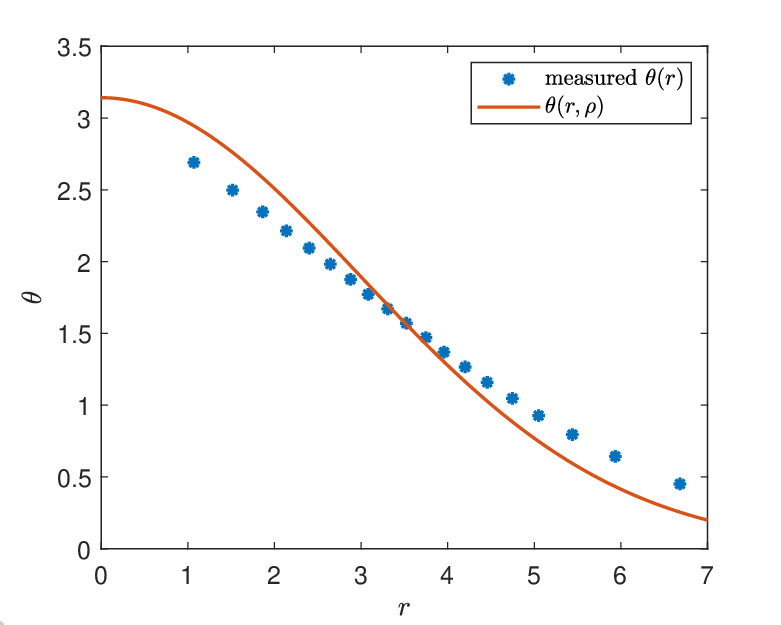}}\\
\subfigure[$d=0.4,b=0.1$]{\includegraphics[width=0.48\textwidth]{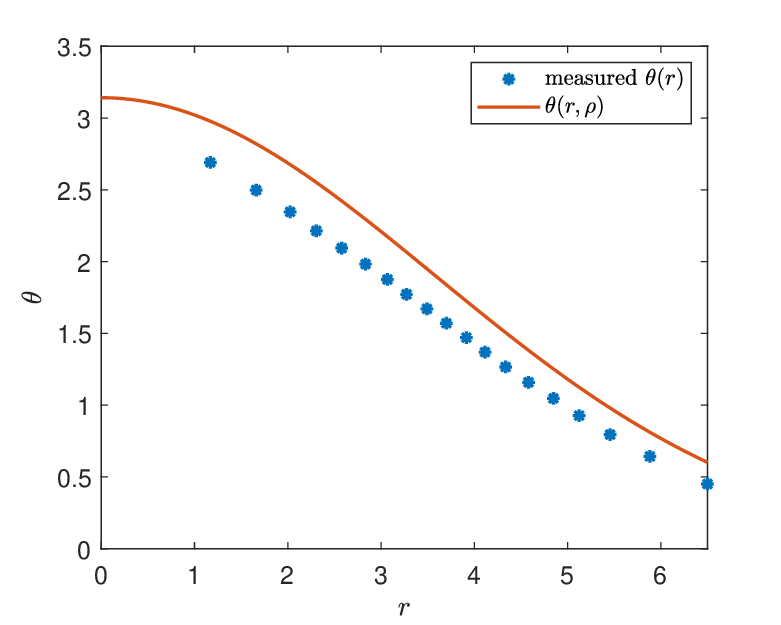}}
\subfigure[$d=0.2,b=0.0025$]{\includegraphics[width=0.48\textwidth]{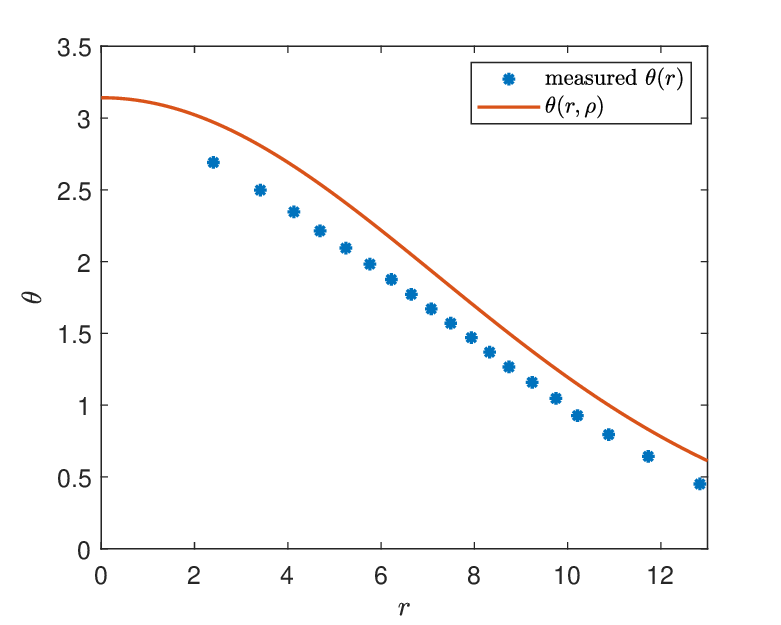}}
  \caption{\label{trishape.fig}The shape of isolated skyrmions in the saturated state}
\end{figure}

The function $\theta (r)$ is often used to describe the shape of a skyrmion~\cite{radius2}.
It has been assumed that the higher order corrections to $\theta (r)$ is small, which in fact requires that the shape of a skyrmion is not changed significantly in the skyrmion phase.
Choosing the configurations at $d = 0.2, b = 0.025$, $d = 0.35, b = 0.1$, $d = 0.4, b = 0.1$ and $d = 0.45, b = 0.15$ as examples, we measure the average $\theta(r)$ with $\theta$ in the range $\cos^{-1}(0.9)\leq \theta \leq \cos^{-1}(-0.9)$.
The results are compared with $\theta_{\rm LO}(r)$ with $r$ rescaled according to $r\to r/\hat{r}_f(s)$, for example, in the case of $d=0.4,b=0.1$, one has $s\approx 0.116165$ and $\theta (r,\rho)=\theta_{\rm LO}(r/0.783516)$.
The results are shown in Fig.~\ref{trishape.fig}.

As shown in Fig.~\ref{trishape.fig}, the shape of isolated skyrmions in a saturated state is similar to the shape of a single isolated skyrmion with $r$ rescaled.
This result indicates that our assumption is valid.
Note that $\theta (r,\rho)$ is also the function $\theta _{\rm LO}(r)$ with $b$ rescaled as $b/\hat{r}$.
It implies that the skyrmions can be seen as being experiencing an effective magnetic strength $B'=B/\hat{r}$ when affected by other skyrmions.

\subsection{\label{level4.4}Matching}

In the calculations and lattice simulations, we use dimensionless parameters.
The numerical results can be matched to the real material by using the rescaling introduced in Ref.~\cite{otherlatticesimulation,match}.
The rescaling factor is denoted as $s$ and $s=(D/J)\lambda /(2\pi \sqrt{2}a)$ where $\lambda$ is helical wavelength and $a$ is the lattice spacing.
The helical wavelength of real materials can be found in Ref.~\cite{review2}.
For example, if we take $\lambda  \approx 60 \;{\rm nm} $, $a=0.4\;{\rm nm}$ and $D/J=0.4$, then $s\approx 6.75$.
Then $r=6.2854$ corresponds to $r=6.2854\times s \times a \approx 16.97\;{\rm nm}$.
Meanwhile the time unit is rescaled as $t'=s^2 J\hbar / J'$, where $J$ is the dimensionless exchange strength and $J'$ is the exchange strength of a real material.
If we choose $J'\approx 3\;{\rm meV}$, the time unit is $t'\approx 0.01\;{\rm ns}$.
The time step in the simulation is $\Delta t=0.01t'\approx 1\;{\rm ps}$.

\section{\label{level5}Summary}

One of the reasons that the skyrmion is proposed as a candidate for the future memory devices is because the size of a skyrmion is small.
The radius of a skyrmion when the density of skyrmions is between the skyrmion lattice and the single isolated skyrmion is an important issue which is lack of exploration.
In this paper, we study the average radius of skyrmions when the density is between a skyrmion lattice and a single isolated skyrmion.
By using the harmonic oscillator expansion, the dependency of the average radius of skyrmions on the parameters of materials, strength of external magnetic field and the density of skyrmions is obtained theoretically.
Then, a lattice simulation of LLG equation is performed to verify our results.

The theoretical result is presented in Eq.~(\ref{eq.4.3}).
Our result indicates that the average radius of skyrmions will decrease with the growth of density even when $b$ and $d$ are unchanged.
The average radii at different $d$ and $b$ are measured by using lattice simulation.
We confirm that our theoretical result can fit the simulated results well.
With this relation, the skyrmion radius for different materials at different densities can be easily predicted.
We also find that, the shapes of the skyrmions are insensitive to the density, which implies that the interactions between skyrmions can be seen as an effective magnetic strength.

\begin{acknowledgments}
This work was partially supported by the National Natural Science Foundation of China under Grants No. 12047570 and the Natural Science Foundation of the Liaoning Scientific Committee No. 2019-BS-154.
\end{acknowledgments}

\bibliography{SkyrmionRadius}
\bibliographystyle{ws-mplb}

\end{document}